# Linearity and Complements in Projective Space

Michael Braun[a], Tuvi Etzion[b,1,*], Alexander Vardy[c,1]

[a] *Faculty of Computer Science, University of Applied Sciences Darmstadt, D-64295 Darmstadt, Germany*
[b] *Department of Computer Science, Technion, Haifa 32000, Israel*
[c] *Department of Electrical Engineering, University of California San Diego, La Jolla, CA 92093, USA*

**Abstract**

The projective space of order $n$ over the finite field $\mathbb{F}_q$, denoted here as $\mathbb{P}_q(n)$, is the set of all subspaces of the vector space $\mathbb{F}_q^n$. The projective space can be endowed with distance function $d_S(X, Y) = \dim(X) + \dim(Y) - 2\dim(X \cap Y)$ which turns $\mathbb{P}_q(n)$ into a metric space. With this, *an $(n, M, d)$ code $\mathbb{C}$ in projective space* is a subset of $\mathbb{P}_q(n)$ of size $M$ such that the distance between any two codewords (subspaces) is at least $d$. Koetter and Kschischang recently showed that codes in projective space are precisely what is needed for error-correction in networks: an $(n, M, d)$ code can correct $t$ packet errors and $\rho$ packet erasures introduced (adversarially) anywhere in the network as long as $2t + 2\rho < d$. This motivates new interest in such codes.

In this paper, we examine the two fundamental concepts of "complements" and "linear codes" in the context of $\mathbb{P}_q(n)$. These turn out to be considerably more involved than their classical counterparts. These concepts are examined from two different points of view, coding theory and lattice theory. Our discussion reveals some surprised phenomena of these concepts in $\mathbb{P}_q(n)$ and leaves some interesting problems for further research.

*Keywords:* network coding, subspace coding, complements, linear codes

## 1. Introduction

Let $\mathbb{F}_q^n$ be the canonical vector space of dimension $n$ over the finite field $\mathbb{F}_q$ of order $q$, where $q$ is a prime power. The *projective space* of order $n$ over $\mathbb{F}_q$, denoted herein by $\mathbb{P}_q(n)$, is the set of all the subspaces of $\mathbb{F}_q^n$, including $\{\mathbf{0}\}$ and $\mathbb{F}_q^n$ itself. Given a nonnegative integer $k \leq n$, the set of all subspaces of $\mathbb{F}_q^n$ that have dimension $k$ is known as a *Grassmannian*, and usually denoted by $\mathbb{G}_q(n, k)$. Thus $\mathbb{P}_q(n) = \cup_{0 \leq k \leq n} \mathbb{G}_q(n, k)$. It is well known that

$$|\mathbb{G}_q(n,k)| = \begin{bmatrix} n \\ k \end{bmatrix} := \frac{(q^n-1)(q^{n-1}-1)\cdots(q^{n-k+1}-1)}{(q^k-1)(q^{k-1}-1)\cdots(q-1)}$$

*Corresponding Author
Email addresses: michael.braun@h-da.de (Michael Braun), etzion@cs.technion.ac.il (Tuvi Etzion), avardy@ucsd.edu (Alexander Vardy)
[1]This research was supported in part by the United States — Israel Binational Science Foundation (BSF), Jerusalem, Israel, under Grant 2006097.

*Preprint submitted to Linear Algebra and Its Applications*        *October 23, 2018*

where $\begin{bmatrix}n\\k\end{bmatrix}$ is the *q-ary Gaussian coefficient*. It turns out that the natural measure of distance, the *subspace distance* in $\mathbb{P}_q(n)$, is given by

$$d_S(X, Y) := \dim(X) + \dim(Y) - 2\dim(X \cap Y) \qquad (1)$$

for all $X, Y \in \mathbb{P}_q(n)$. It is well known (cf. [1, 10]) that the function above is a metric; thus both $\mathbb{P}_q(n)$ and $\mathbb{G}_q(n, k)$ can be regarded as metric spaces. Given a metric space, one can define codes. We say that $\mathbb{C} \subseteq \mathbb{P}_q(n)$ is an $(n, M, d)$ *code in projective space* if $|\mathbb{C}| = M$ and $d_S(X, Y) \geq d$ for all $X, Y \in \mathbb{C}$. If an $(n, M, d)$ code $\mathbb{C}$ is contained in $\mathbb{G}_q(n, k)$ for some $k$, we say that $\mathbb{C}$ is an $(n, M, d, k)$ code. An $(n, M, d, k)$ code is also called a *constant dimension code*.

The $(n, M, d)$, respectively $(n, M, d, k)$, codes in projective space are akin to the familiar codes in the Hamming space, respectively (constant weight) codes in the Johnson space, where the Hamming distance serves as the metric. There are, however, important differences. For all $q, n$ and $k$, the metric space $\mathbb{G}_q(n, k)$ corresponds to a distance-regular graph, similar to the distance-regular graph resulting from the Johnson space. On the other hand, while the Hamming space $\mathbb{F}_q^n$ is always distance-regular (as a graph), the projective space $\mathbb{P}_q(n)$ is not. This implies that conventional geometric intuition does not always apply.

Codes in $\mathbb{G}_q(n, k)$ were studied, somewhat sparsely, over the past twenty years. For example, the nonexistence of perfect codes in $\mathbb{G}_q(n, k)$ was proved in [3] and again in [12]. In [1] it was shown that "Steiner structures" yield diameter-perfect codes in $\mathbb{G}_q(n, k)$; properties of these structures were studied in [14]. It appears that codes in the projective space $\mathbb{P}_q(n)$ were not studied at all, until recently, e.g. [8, 9, 10, 11, 16, 18].

Recently, Koetter and Kschischang [10] showed that codes in $\mathbb{P}_q(n)$ are precisely what is needed for error-correction in networks: an $(n, M, d)$ code can correct any $t$ packet errors and any $\rho$ packet erasures introduced (adversarially) anywhere in the network as long as $2t + 2\rho < d$. This motivates our interest in such codes.

The well known concept of $q$-analogs replaces subsets by subspaces of a vector space over a finite field and their sizes by dimensions of the subspaces. In this respect, constant dimension codes are the $q$-analog of constant weight codes.

The goal of this paper is to examine two basic concepts in coding theory, namely "complements" and "linear codes". These concepts are well-known in coding theory for binary codes and codes over $\mathbb{F}_q$, respectively. Various problems concerning complements of subspaces over $\mathbb{F}_q$ were considered in the past, e.g. [4, 5, 6]. Our goal is to discuss these concepts in the projective space. We will tackle these concepts from two different points of view, coding theory and lattice theory.

The rest of this paper is organized as follows. In Section 2 we will give a formal definition for the term complement. Four properties will be required. A function which satisfies only some of these properties will be called quasi-complement. We will consider each of the fifteen nonempty subsets of these four properties for the existence of quasi-complements. We will discuss what is the largest subset of $\mathbb{P}_q(n)$ on which a complement function can be defined. In Section 3 we define linear and quasi-linear codes in $\mathbb{P}_q(n)$ and show examples of linear codes of small size. We conjecture and give some evidence that larger codes might not exist.

In Section 4 we present our concepts through another point of view, lattice theory. We prove some connection between properties of lattices and quasi-complements. Open problems for further research are presented in Section 5.

## 2. Complements in Projective Space

The concept of complement in the Hamming space is rather simple. It applies only for binary words. There is no definition for complements over $\mathbb{F}_q$. In $\mathbb{P}_q(n)$ we will consider any $q \geq 2$ for complements for most of our discussion. Two binary words $x = (x_1, \ldots, x_n)$ and $y = (y_1, \ldots, y_n)$ are complements if for each $i$, $1 \leq i \leq n$, we have $x_i + y_i = 1$, i.e., $x_i = 0$ if and only if $y_i = 1$. This definition implies a few properties for the complements. One of them is that the bit by bit addition of two words which are complements is the all-ones word. This motivates to define the complement of a subspace $X \in \mathbb{P}_q(n)$ as a subspace $Y \in \mathbb{P}_q(n)$ such that their sum is direct and yields the complete vector space, i.e. $X \oplus Y = \mathbb{F}_q^n$. The main problem in this definition is that $Y$ is not unique.

Therefore, let us consider the key properties of the complement mapping $f : x \mapsto \overline{x} = \mathbf{1} + x$ in the Hamming space, then translate these properties into the context of $\mathbb{P}_q(n)$. For a binary word $x = (x_1, \ldots, x_n)$, let $\text{supp}(x)$ denote the *support* of $x$, i.e. the set of coordinates in $x$ which are equal to 1, $\text{supp}(x) = \{i \ : \ x_i = 1, \ 1 \leq i \leq n\}$. Let $\text{wt}(x)$ denote the *weight* of $x$, i.e. the number of coordinates in $x$ which are equal to 1, $\text{wt}(x) = |\text{supp}(x)|$. The most basic property of complements in $\mathbb{F}_q^n$ is this: for all $x \in \mathbb{F}_q^n$, we have $\text{supp}(x) \cap \text{supp}(\overline{x}) = \varnothing$ and $\text{supp}(x) \cup \text{supp}(\overline{x}) = \{1, \ldots, n\}$. In other words, for all $x \in \mathbb{F}_q^n$, the support of $\overline{x}$ is the complement of the support of $x$ with respect to the universal set $\{1, \ldots, n\}$. This is why $f$ is, indeed, called the "complement". This property implies that if $\text{wt}(x) = k$ then $\text{wt}(\overline{x}) = n - k$, for all $x \in \mathbb{F}_q^n$. Moreover, the complement mapping $f$ is a bijection between the set of vectors of weight $k$ and the set of vectors of weight $n - k$ in $\mathbb{F}_q^n$. Another key property is that $f$ is an involution: the complement of the complement of $x$ is $x$ itself, for all $x \in \mathbb{F}_q^n$. Finally, the most useful feature of $f$, in coding theory, is that it is distance preserving (isometric). Namely, for all $x, y \in \mathbb{F}_q^n$, we have $d_H(\overline{x}, \overline{y}) = d_H(x, y)$, where $d_H(\cdot, \cdot)$ is the Hamming distance. All this leads to the following definition.

**Definition 1.** Let $\mathcal{U}$ be a subset of $\mathbb{P}_q(n)$ and let $\mathcal{U}_k := \mathbb{G}_q(n,k)|_\mathcal{U}$ be the set of all $k$-subspaces in $\mathcal{U}$. We say that a function $f : \mathcal{U} \to \mathcal{U}$ is a *complement on* $\mathcal{U}$ (and denote $\overline{X} = f(X)$ for all $X \in \mathcal{U}$) if $f$ has the following properties:

**P1.** $X \cap \overline{X} = \{\mathbf{0}\}$ and $X + \overline{X} = \mathbb{F}_q^n$, i.e. $X \oplus \overline{X} = \mathbb{F}_q^n$ for all $X \in \mathcal{U}$.
**P2.** $f$ establishes a bijection between $\mathcal{U}_k$ and $\mathcal{U}_{n-k}$ for all $k$, $0 \leq k \leq n$.
**P3.** $f(f(X)) = X$ for all $X \in \mathcal{U}$.
**P4.** $d_S(\overline{X}, \overline{Y}) = d_S(X, Y)$ for all $X, Y \in \mathcal{U}$.

Unfortunately, as we shall see, complements do not always exist. Thus the following terminology is useful. Given a function $f : \mathcal{U} \to \mathcal{U}$ that has some of the properties **P1** – **P4** in Definition 1, but not others, we say that $f$ is a *quasi-complement on* $\mathcal{U}$ (and indicate which properties $f$ has).

A natural candidate for a complement function on the entire projective space $\mathbb{P}_q(n)$ is the *orthogonal complement* over $\mathbb{F}_q$ (usually known as the *dual code* in coding theory), given by $X^\perp = \{y \in \mathbb{F}_q^n : \langle x, y \rangle = 0 \text{ for all } x \in X\}$, where $\langle \cdot, \cdot \rangle$ is the standard inner product over $\mathbb{F}_q$. However, this turns out to be only a quasi-complement, since property **P1** is not satisfied. Nevertheless, the orthogonal complement is very important in our discussion. For a code $\mathbb{C}$ in $\mathbb{P}_q(n)$, the *orthogonal complement code* of $\mathbb{C}$ is defined by $\mathbb{C}^\perp = \{X^\perp : X \in \mathbb{C}\}$. The following lemma was proved in [10].

**Lemma 1.** *If $X$ and $Y$ be two subspaces of $\mathbb{F}_q^n$ then $d_S(X^\perp, Y^\perp) = d_S(X, Y)$.*

**Theorem 1.** *Let $f : \mathbb{P}_q(n) \to \mathbb{P}_q(n)$ be the orthogonal complement function defined by $f(X) = X^\perp$. Then $f$ is a quasi-complement on $\mathbb{P}_q(n)$, satisfying properties **P2**, **P3**, and **P4**.*

PROOF. Properties **P2** and **P3** follow directly from the definitions and **P4** follows from Lemma 1. □

Since **P1** is, arguably, the most salient property of a complement, it is tempting to construct complements as follows. Given $X \in \mathbb{P}_q(n)$, find a subspace $Y$ such that $X \cap Y = \{\mathbf{0}\}$ and $X + Y = \mathbb{F}_q^n$ (say, by extending a basis for $X$ to a basis for $\mathbb{F}_q^n$), and let $Y$ be the complement of $X$. But the problem is that such a subspace $Y$ is not unique. If one selects $Y$ arbitrarily from the multitude of possibilities, property **P2** could be violated. Without **P2**, complementary codes $\mathbb{C} \subseteq \mathbb{P}_q(n)$ and $\overline{\mathbb{C}} = \{\overline{X} : X \in \mathbb{C}\}$ could have different sizes, which is clearly undesirable. Can one construct $X$ in some "canonical" manner, so that both **P1** and **P2** are satisfied? For a simple answer we need the following definition and well known lemma [17].

An *one-factor* in an undirected graph is a set of edges $\mathcal{E}$ such that each vertex in the graph is adjacent to exactly one edge in $\mathcal{E}$.

**Lemma 2.** *A regular bipartite graph with two equal sides has an one-factor.*

**Theorem 2.** *For all positive $n$, there exists a quasi-complement $f : \mathbb{P}_q(n) \to \mathbb{P}_q(n)$ that satisfies properties **P1** and **P2**.*

PROOF. For each $k \leq \frac{n}{2}$ we construct a graph $G(k)$ which has two set of vertices $\mathcal{V}_k$ and $\mathcal{W}_k$. $\mathcal{V}_k$ consists of the vertices of $\mathbb{G}_q(n, k)$ and $\mathcal{W}_k$ consists of the vertices of $\mathbb{G}_q(n, n-k)$. A vertex $X \in \mathcal{V}_k$ is connected to a vertex $Y \in \mathcal{W}_k$ if $X \cap Y = \{\mathbf{0}\}$. Clearly, $|\mathcal{V}_k| = |\mathcal{W}_k|$, and the degree of a vertex in $\mathcal{V}_k$ is the number of $(n-k)$-dimensional subspaces which are disjoint with a given $k$-dimensional subspace, and vice-versa. Hence, the graph is a regular bipartite graph with two equal sizes. By Lemma 2, $G(k)$ has an one-factor $\mathcal{F}$. If the edge which connects $X \in \mathcal{V}_k$ and $Y \in \mathcal{W}_k$ is in $\mathcal{F}$ then we set $Y = f(X)$ and $X = f(Y)$. It is easily verified that $f$ satisfies **P1** and **P2**. □

The proof of Theorem 2 is useful in the following result.

**Theorem 3.** *A quasi-complement* $f : \mathbb{P}_q(n) \to \mathbb{P}_q(n)$ *that satisfies properties* **P1**, **P2**, *and* **P3** *exists if and only if $n$ is odd or $q$ is odd.*

PROOF. If $n$ is odd then the function $f$ from the proof of Theorem 2 satisfies also **P3**. For $n = 2k$ we have that $\begin{bmatrix} n \\ k \end{bmatrix}$ is an even integer if and only if $q$ is odd. Hence for a given $f : \mathbb{P}_q(n) \to \mathbb{P}_q(n)$, $q$ even, either **P3** is not satisfied or there exist at least one $X \in \mathbb{G}_q(2k, k)$ for which $f(X) = X$ and **P1** is not satisfied. For $q$ odd, it was proved in [4] that the graph consisting of the vertices of $\mathbb{G}_q(2k, k)$, where two vertices $X$ and $Y$ are connected if $X \cap Y = \{\mathbf{0}\}$, has an one-factor. This implies the required result. □

Similarly, we can prove

**Theorem 4.** *There exists a quasi-complement* $f : \mathbb{P}_q(n) \to \mathbb{P}_q(n)$ *that satisfies properties* **P1** *and* **P3** *if and only if $q$ is odd or $n$ is odd.*

**Remark 1.** The results concerning Theorems 2, 3, and 4 appears in [4]. We gave their proof for completeness.

Do complements in projective space, at all, exist? The answer is yes, but on sets smaller than the whole of $\mathbb{P}_q(n)$. The following theorem exhibits a large subset of $\mathbb{P}_q(n)$ on which a complement can be defined.

**Theorem 5.** *Let* $\mathcal{V}_q(n) := \{X \in \mathbb{P}_q(n) : X \cap X^\perp = \{\mathbf{0}\}\}$ *and define the mapping* $f : \mathcal{V}_q(n) \to \mathcal{V}_q(n)$ *by* $f(X) = X^\perp$. *Then $f$ is a complement on $\mathcal{V}_q(n)$.*

PROOF. By Theorem 1, $f$ satisfies **P2**, **P3**, and **P4**. If $X \cap X^\perp = \{\mathbf{0}\}$, for each $X \in \mathbb{F}_q$, and **P2** is satisfied, then $X \oplus X^\perp = \mathbb{F}_q^n$ and hence **P1** is also satisfied. □

A closed-form expression for $|\mathcal{V}_q(n)|$ was given by Sendrier [15]. Using the results of [15], it can be shown that the size of $\mathcal{V}_q(n)$ is *proportional* to $|\mathbb{P}_q(n)|$, specifically:

$$\lim_{n \to \infty} \frac{|\mathcal{V}_q(n)|}{|\mathbb{P}_q(n)|} = \prod_{i=1}^{\infty} \frac{1}{1 + q^{-i}} \ .$$

The limit converge to $0.4194\ldots$ when $q = 2$, $0.639\ldots$ when $q = 3$, $0.7375\ldots$ when $q = 4$, and $0.9961\ldots$ when $q = 256$.

We conjecture that $\mathcal{V}_q(n)$ is, in fact, the largest subset of $\mathbb{P}_q(n)$ on which a complement can be defined. Some evidence for this conjecture is provided by the observation that $\mathcal{V}_2(n)$ attains the bound of Proposition 1, which follows, with equality, as implied by the following two lemmas whose proofs is immediate.

**Lemma 3.** *If $X$ is an one-dimensional subspace of $\mathbb{P}_2(n)$ then $X \cap X^\perp = \{\mathbf{0}\}$ if and only if the number of the nonzero elements contained in $X$ is odd.*

PROOF. Follows immediately from the fact that $X \subset X^\perp$ if and only if the number of the nonzero elements contained in $X$ is even. □

**Lemma 4.** *If $Y_1, Y_2, Y_3$ are three distinct $(n-1)$-dimensional subspaces of $\mathbb{F}_2^n$ such that $\dim(Y_1 \cap Y_2 \cap Y_3) = n - 2$ then $Y_1 \cup Y_2 \cup Y_3 = \mathbb{F}_2^n$.*

PROOF. Let $V = Y_1 \cap Y_2 \cap Y_3$ and $Z_i = Y_i \setminus V$ for $i = 1, 2, 3$. Clearly, $Z_i \cap Z_j = \varnothing$ for $i \neq j$, and $|Z_i| = 2^{n-2}$. Hence, $|V| + |Z_1| + |Z_2| + |Z_3| = 2^n$ and therefore $Y_1 \cup Y_2 \cup Y_3 = \mathbb{F}_2^n$. □

**Lemma 5.** *Let $\mathcal{U}$ be a subset of $\mathbb{P}_2(n)$, and suppose that there exists a complement on $\mathcal{U}$. If $X$, $Y$, and $Z$ are three distinct $(n-1)$-dimensional subspaces of $\mathcal{U}$ then $\dim(X \cap Y \cap Z) = n - 3$.*

PROOF. Assume the contrary, that there exist three distinct $(n-1)$-dimensional subspaces $Y_1$, $Y_2$, and $Y_3$ in $\mathcal{U}$, such that $\dim(Y_1 \cap Y_2 \cap Y_3) = n - 2$. Let $T = Y_1 \cap Y_2 \cap Y_3$ and let $P_i = Y_i \setminus T$, $i = 1, 2, 3$. Let $f(Y_i) = X_i$, $i = 1, 2, 3$, where $X_i$ is an one-dimensional subspace of $\mathcal{U}$. By **P1** we have that $X_i \cap Y_i = \{\mathbf{0}\}$ and by **P3** we have $f(X_i) = Y_i$, $i = 1, 2, 3$. Hence, as a consequence of Lemma 4, we can assume w.l.o.g. that $X_1 \setminus \{\mathbf{0}\} \subset P_2$ (otherwise, $X_1 \setminus \{\mathbf{0}\} \subset P_3$), i.e., $d_S(X_1, Y_2) = n - 2$. Similarly, either $X_2 \setminus \{\mathbf{0}\} \subset P_3$ or $X_2 \setminus \{\mathbf{0}\} \subset P_1$. We distinguish between these two cases:
**Case 1:** If $X_2 \setminus \{\mathbf{0}\} \subset P_3$ then $X_2 \cap Y_1 = \{\mathbf{0}\}$. Hence, $d_S(f(X_1), f(Y_2)) = d_S(Y_1, X_2) = n$. But, $d_S(X_1, Y_2) = n - 2$, a contradiction to **P4**.
**Case 2:** Assume that $X_2 \setminus \{\mathbf{0}\} \subset P_1$. Again, as a consequence of Lemma 4, we have either $X_3 \setminus \{\mathbf{0}\} \subset P_1$ (and hence $X_3 \cap Y_2 = \{\mathbf{0}\}$) or $X_3 \setminus \{\mathbf{0}\} \subset P_2$ (and hence $X_3 \cap Y_1 = \{\mathbf{0}\}$). Again, we distinguish between these two cases:
**Case 2.1:** If $X_3 \setminus \{\mathbf{0}\} \subset P_1$ then $d_S(X_3, Y_1) = n - 2$. Hence, by **P4** we have $n - 2 = d_S(X_3, Y_1) = d_S(f(Y_3), f(X_1)) = d_S(Y_3, X_1)$, i.e., $X_1 \setminus \{\mathbf{0}\} \subset P_3$, a contradiction since $X_1 \setminus \{\mathbf{0}\} \subset P_2$ and $P_2 \cap P_3 = \varnothing$.
**Case 2.2:** If $X_3 \setminus \{\mathbf{0}\} \subset P_2$ then $d_S(X_3, Y_2) = n - 2$. Hence, by **P4** we have $n - 2 = d_S(X_3, Y_2) = d_S(f(Y_3), f(X_2)) = d_S(Y_3, X_2)$, i.e., $X_2 \setminus \{\mathbf{0}\} \subset P_3$, a contradiction since $X_2 \setminus \{\mathbf{0}\} \subset P_1$ and $P_1 \cap P_3 = \varnothing$.

Thus, we proved that for any three distinct $(n-1)$-dimensional subspaces $X$, $Y$, $Z$ of $\mathcal{U}$ we have $\dim(X \cap Y \cap Z) = n - 3$. □

**Proposition 1.** *Let $\mathcal{U}$ be a subset of $\mathbb{P}_2(n)$, and suppose that there exists a complement on $\mathcal{U}$. Then $\mathcal{U}$ contains at most $2^{n-1}$ one-dimensional subspaces of the $2^n - 1$ one-dimensional subspaces of $\mathbb{F}_2^n$.*

PROOF. Assume $f : \mathcal{U} \to \mathcal{U}$ is a function satisfying properties **P1** – **P4**. If $X$ and $Y$ are two distinct $(n-1)$-dimensional subspaces of $\mathbb{F}_q^n$ then clearly $\dim(X \cap Y) = n - 2$.

Let $Y$ be an $(n-1)$-dimensional subspace of $\mathcal{U}$. $Y$ contains $2^{n-1} - 1$ distinct $(n-2)$-dimensional subspaces. By Lemma 5, each such $(n-2)$-dimensional subspace can be a subspace of at most another one $(n-1)$-dimensional subspace of $\mathcal{U}$, except for $Y$. Since $Y$ intersect each $(n-1)$-dimensional subspace of $\mathcal{U}$ in an $(n-2)$-dimensional subspace, it follows that $\mathcal{U}$ contains at most $2^{n-1} - 1$ subspaces of dimension $n - 1$, except for $Y$. Thus, $\mathcal{U}$ contains at most $2^{n-1}$ one-dimensional subspaces. □

We have considered quasi-complements for all subsets of properties, **P1**, **P2**, **P3**, and **P4**, except for quasi-complements which satisfy properties **P1** and **P4**. These quasi-complements will be discussed in Section 4.

## 3. Linear Codes in Projective Space

The major part of classical coding theory is concerned with *linear codes*. Linear codes have a succinct representation in terms of generator and parity-check matrices, and possess numerous other properties that greatly facilitate both construction and decoding of such codes. Therefore, it is natural to ask whether there exist linear codes in projective space. While the question itself is simple, answering it turns out to be quite complicated. In order to simplify the matters somewhat, we shall restrict our attention in this section to binary codes — namely, codes in the projective space $\mathbb{P}_2(n)$ over $\mathbb{F}_2$ (although some of our results extend straightforwardly to arbitrary finite fields).

Let us begin by elucidating the key properties of linearity in the Hamming space $\mathbb{F}_2^n$, and using them to define clearly what we mean by linearity in $\mathbb{P}_2(n)$. Evidently, the most basic property of any linear code $C \subseteq \mathbb{F}_2^n$ is that $C$ is a vector space over $\mathbb{F}_2$. What makes such codes possible is the innocuous operation of componentwise addition modulo 2, which turns the set of $n$-tuples over $\mathbb{F}_2$ into a vector space. The identity element $\mathbf{0}$ of this addition is, of course, the all-zero $n$-tuple. This leads, by analogy, to the following definition.

**Definition 2.** Let $\mathcal{U}$ be a subset of $\mathbb{P}_2(n)$ with $\{\mathbf{0}\} \in \mathcal{U}$. We say that $\mathcal{U}$ is a *quasi-linear code* in $\mathbb{P}_2(n)$ if there exists a function $\boxplus : \mathcal{U} \times \mathcal{U} \to \mathcal{U}$ such that $(\mathcal{U}, \boxplus)$ is an abelian group with the following properties: the identity element is $\{\mathbf{0}\}$, and the inverse of every group element $X \in \mathcal{U}$ is $X$ itself. The function $\boxplus$ is then said to be a *linear addition* on $\mathcal{U}$.

Some remarks on this definition appear to be in order. Why do we require that the inverse of every $X \in \mathcal{U}$ is $X$ itself? This is precisely what turns the abelian group $(\mathcal{U}, \boxplus)$ into a "vector space" over $\mathbb{F}_2$. The standard and, essentially, the only way to define scalar multiplication over $\mathbb{F}_2 = \{0, 1\}$ is this: $1X = X$ and $0X = \{\mathbf{0}\}$ for all $X \in \mathcal{U}$. Distributivity holds for this scalar multiplication over addition in $\mathbb{F}_2$ if and only if every $X \in \mathcal{U}$ is its own inverse, since:
$$\{\mathbf{0}\} = 0X = (1+1)X = 1X \boxplus 1X = X \boxplus X$$
Why do we explicitly require that the identity element of $\boxplus$ is the null-space $\{\mathbf{0}\}$? Doesn't this follow from the other parts of the definition? The answer is that it does not. Suppose that $\mathcal{U}$ is a quasi-linear code in $\mathbb{P}_2(n)$ with the identity element $\{\mathbf{0}\}$ and linear addition $\boxplus$. Let $\mathcal{U}' = \{X^\perp : X \in \mathcal{U}\}$, where $X^\perp$ is the usual orthogonal complement of $X \in \mathbb{F}_2^n$. Define the addition $\boxplus'$ on $\mathcal{U}'$ by $X^\perp \boxplus' Y^\perp := (X \boxplus Y)^\perp$. Then $(\mathcal{U}', \boxplus')$ is an abelian group and every element of $\mathcal{U}'$ is its own inverse. However, the identity element of $(\mathcal{U}', \boxplus)$ is a strange one: it is the ambient vector space $\mathbb{F}_2^n = \{\mathbf{0}\}^\perp$ rather than $\{\mathbf{0}\}$ (since $X \boxplus' \mathbb{F}_2^n = (X^\perp \boxplus \{\mathbf{0}\})^\perp = (X^\perp)^\perp = X$). Further justification for using the null-space as the identity element will be given in Section 4. Finally, since a quasi-linear code $\mathcal{U}$ is a vector space over $\mathbb{F}_2$, why do we say that it is "quasi-linear" rather than "linear"? The answer is provided in the following proposition.

**Proposition 2.** *Let $\mathcal{U}$ be a subset of $\mathbb{P}_2(n)$ with $\{\mathbf{0}\} \in \mathcal{U}$. Then $\mathcal{U}$ is a quasi-linear code if and only if $|\mathcal{U}|$ is a power of 2.*

PROOF. Since a quasi-linear code in $\mathbb{P}_2(n)$ is a vector space over $\mathbb{F}_2$, the condition that $|\mathcal{U}| = 2^m$ for some integer $m$ is trivially necessary for $\mathcal{U}$ to be quasi-linear. To see that it is also sufficient, establish an *arbitrary bijection* $g$ between $\mathcal{U}$ and $\mathbb{F}_2^m$ (that maps $\{\mathbf{0}\}$ to $\mathbf{0}$), and define the linear addition $\boxplus$ on $\mathcal{U}$ by $X \boxplus Y := g^{-1}(g(X) + g(Y))$, where $+$ is the addition in $\mathbb{F}_2^m$. $\square$

**Remark 2.** If we do not require that the null-space is the identity element then large linear codes in $\mathbb{P}_2(n)$ do exist. Let $k$ be an integer, $0 < k < n$, and let $I_k$ be an identity matrix of order $k$. Define the set of $k \times n$ matrices

$$M := \{[I_k \mid A] : A \in \mathbb{F}_2^{k \times (n-k)}\}.$$

For $[I_k \mid A_1], [I_k \mid A_2] \in M$ we define $[I_k \mid A_1] + [I_k \mid A_2] = [I_k \mid A_1 + A_2]$. The rows of the matrix $[I_k \mid A]$ can be viewed as a basis of a $k$-dimensional subspace in $\mathbb{P}_2(n)$. Let $P(M)$ be the set of $k$-dimensional subspaces defined by $M$. The addition defined makes $M$ to be a linear code in $\mathbb{P}_2(n)$ whose size is $2^{k(n-k)}$ with the $k$-dimensional subspace defined by the matrix $[I_k \mid \mathbf{0}]$ being the identity element. One can use a permutation $\pi$ on $n$ elements and apply it on the columns of $M$ to obtain a different linear code in $\mathbb{P}_2(n)$. It is also easy to verify that the construction is generalized for $\mathbb{P}_q(n)$. Error-correcting codes in $\mathbb{P}_q(n)$ based on this construction of linear codes were defined in [7, 16].

Proposition 2 clearly shows that the property of being a vector space over $\mathbb{F}_2$ is not, in itself, sufficient to impose useful structure. What makes linear codes in the Hamming space so useful in coding theory is the fact that the vector-space addition (componentwise modulo 2) preserves distances. Geometrically, a linear code "looks the same" from any codeword, and its distance distribution coincides with its weight distribution. This leads to the following definition.

**Definition 3.** *Let $\mathcal{U}$ be a subset of $\mathbb{P}_2(n)$ with $\{\mathbf{0}\} \in \mathcal{U}$. We say that $\mathcal{U}$ is a linear code in $\mathbb{P}_2(n)$ if it is quasi-linear and the corresponding linear addition $\boxplus$ is isometric, namely:*

$$d_S(X \boxplus Y_1, X \boxplus Y_2) = d_S(Y_1, Y_2) \quad \text{for all} \quad X, Y_1, Y_2 \in \mathcal{U}$$

It is obvious that if $\mathcal{U}$ is a linear code in $\mathbb{P}_2(n)$, then any subspace of $\mathcal{U}$ (as a vector space over $\mathbb{F}_2$) is also a linear code, for the same addition $\boxplus$. Thus if we can endow a large subset of $\mathbb{P}_2(n)$ with linear structure, we immediately obtain a large "ambient space" wherein linearity is inherent. This leads to the following question: what is the "largest" linear code in $\mathbb{P}_2(n)$? A partial answer to this question is provided in the following theorem. For a set $\mathcal{S} \subseteq \mathbb{F}_2^n$, let $\langle \mathcal{S} \rangle$ denote the subspace of $\mathbb{F}_2^n$ spanned by the elements of $\mathcal{S}$ (with $\langle \varnothing \rangle = \{\mathbf{0}\}$).

**Theorem 6.** *Let $\{e_1, e_2, \ldots, e_n\}$ be an arbitrary basis for $\mathbb{F}_2^n$ over the subfield $\mathbb{F}_2$, and let $\mathcal{L} = \{\langle \mathcal{S} \rangle : \mathcal{S} \subseteq \{e_1, e_2, \ldots, e_n\}\}$. Then $\mathcal{L}$ is a linear code of size $2^n$ in $\mathbb{P}_2(n)$.*

PROOF. Let $\mathcal{B} = \{e_1, e_2, \ldots, e_n\}$. Given an element $X$ of $\mathcal{L}$, let $X|_\mathcal{B} = X \cap \mathcal{B}$. We define an addition $\boxplus$ on $\mathcal{L}$ as follows:

$$X \boxplus Y := \Big\langle (X|_\mathcal{B} \cup Y|_\mathcal{B}) \setminus (X|_\mathcal{B} \cap Y|_\mathcal{B}) \Big\rangle$$

In other words, $X \boxplus Y$ is the linear span of those elements of the basis $\mathcal{B}$ that are contained in either $X$ or $Y$ but not both. It can be verified that $\boxplus$ is an isometric linear addition on $\mathcal{L}$. □

The construction of Theorem 6 effectively embeds a copy of $\mathbb{F}_2^n$ in $\mathbb{P}_2(n)$. Each element $V$ of $\mathcal{L}$ can be identified with the binary vector $x = (x_1, \ldots, x_n)$ such that $x_i = 1$ if and only if $e_i \in (V \cap \mathcal{B})$. Given this bijection, *all* the results for (linear) codes in $\mathbb{F}_q^n$ carry over immediately to projective space; in particular, a subset of $\mathcal{L}$ has minimum distance $d$ in $\mathbb{P}_2(n)$ if and only if $d$ is the minimum Hamming distance of the corresponding subset of $\mathbb{F}_q^n$. This makes it possible to construct codes in $\mathbb{P}_2(n)$ from codes in $\mathbb{F}_2^n$ (and vice versa).

The problem is that the linear code constructed in Theorem 6 is rather small: $|\mathcal{L}| = 2^n$ whereas $|\mathbb{P}_2(n)| > 2^{0.25n^2}$. Are there larger linear codes in $\mathbb{P}_2(n)$? Unfortunately, we believe that the answer is no. Some evidence for this conjecture is provided in what follows.

First, we establish certain conditions that any isometric linear addition has to satisfy.

**Lemma 6.** *Let $\mathcal{U}$ be a linear code in $\mathbb{P}_2(n)$ and let $\boxplus$ be the isometric linear addition on $\mathcal{U}$. Then for all $X, Y \in \mathcal{U}$, we have:*

$$\dim(X \boxplus Y) = \dim(X) + \dim(Y) - 2\dim(X \cap Y)$$

PROOF. By Definitions 2 and 3, we get $d_S(X, Y) = d_S(X \boxplus Y, Y \boxplus Y) = d_S(X \boxplus Y, \{\mathbf{0}\}) = \dim(X \boxplus Y)$. By (1) we have that if $X$ and $Y$ are two subspaces then $d_S(X, Y) = \dim(X) + \dim(Y) - 2\dim(X \cap Y)$ and the lemma follows. □

**Lemma 7.** *For any three subspaces $X$, $Y$, and $Z$ of a linear code $\mathcal{U}$ in $\mathbb{P}_2(n)$ with isometric linear addition $\boxplus$ the condition $Z = X \boxplus Y$ implies $Y = X \boxplus Z$.*

PROOF. If $Z = X \boxplus Y$ then $X \boxplus Z = X \boxplus X \boxplus Y = Y$ by Definition 2. □

**Lemma 8.** *Let $\mathcal{U}$ be a linear code in $\mathbb{P}_2(n)$ and let $\boxplus$ be the isometric linear addition on $\mathcal{U}$. If $X$ and $Y$ are any two elements of $\mathcal{U}$ such that $X \cap Y = \{\mathbf{0}\}$, then $X \boxplus Y = X + Y$, where $+$ denotes the conventional sum of vector spaces.*

PROOF. Let $T = X \boxplus Y$ and assume the contrary that $X \boxplus Y \neq X + Y$. By Lemma 6 we have $\dim(X \boxplus Y) = \dim(X) + \dim(Y)$. Since $\dim(T) = \dim(X \boxplus Y) = \dim(X) + \dim(Y) = \dim(X + Y)$ and $T \neq X + Y$ it follows that $T$ doesn't contain $X$ or $T$ doesn't contain $Y$. W.l.o.g. we assume that $T$ doesn't contain $X$. It implies that $\dim(X) > \dim(T \cap X)$. By Lemma 7 we have that $Y = T \boxplus X$ and therefore by Lemma 6 we have, $\dim(Y) = \dim(T) + \dim(X) - 2\dim(T \cap X) = \dim(X) + \dim(Y) + \dim(X) - 2\dim(T \cap X) > \dim(Y)$, a contradiction. Thus $X \subseteq T$ and similarly $Y \subseteq T$, and therefore, $X \boxplus Y = T = X + Y$. □

Using Lemmas 6 and 8, we can prove the following result, which lends credence to the conjecture that the largest linear code in the projective space $\mathbb{P}_2(n)$ has size $2^n$.

**Proposition 3.** *Let $\mathcal{U}$ be a linear code in $\mathbb{P}_2(n)$ with isometric linear addition $\boxplus$. Then $\mathcal{U}$ contains at most $n$ of the $2^n - 1$ one-dimensional subspaces of $\mathbb{F}_2^n$.*

PROOF. Assume the contrary, i.e., $\mathcal{U}$ is a linear code in $\mathbb{P}_2(n)$ with at least $n+1$ one-dimensional subspaces of $\mathbb{F}_2^n$. Let $\{e_1, e_2, \ldots, e_{m+1}\}$ be the smallest subset of dependent elements from $\mathbb{F}_2^n$ for which each one-dimensional subspace $\mathcal{E}_i = \{0, e_i\}$ is contained in $\mathcal{U}$. Such a subset of dependent elements exists since $\mathcal{U}$ contains at least $n+1$ one-dimensional subspaces of $\mathbb{F}_2^n$. Clearly, any $m$-subset of the set $\{e_1, e_2, \ldots, e_{m+1}\}$ is a set of independent elements. Hence, $\sum_{i=0}^{m+1} e_i = 0$ and $(\mathcal{E}_1 + \mathcal{E}_2 + \cdots + \mathcal{E}_{m-1}) + \mathcal{E}_m = (\mathcal{E}_1 + \mathcal{E}_2 + \cdots + \mathcal{E}_{m-1}) + \mathcal{E}_{m+1}$ (since the addition of any $m$ subspaces contains the $(m+1)$-th subspace). This implies by repeatedly using Lemma 8 that $(\mathcal{E}_1 \boxplus \mathcal{E}_2 \boxplus \cdots \boxplus \mathcal{E}_{m-1}) \boxplus \mathcal{E}_m = (\mathcal{E}_1 \boxplus \mathcal{E}_2 \boxplus \cdots \boxplus \mathcal{E}_{m-1}) \boxplus \mathcal{E}_{m+1}$, a contradiction to the fact that $(\mathcal{U}, \boxplus)$ is a group by Definition 2. □

On the other hand, we can also show that the linear code $\mathcal{L}$ of size $2^n$ constructed in Theorem 6 is *not* unique. For $n \geq 3$, there exist linear codes of the same size in $\mathbb{P}_2(n)$ whose dimension distribution is different from that of $\mathcal{L}$. First, we construct such a code $\Psi$ for $n = 3$.

**Example 1.** $\Psi$ consists of the null-space and the seven two-dimensional subspaces of $\mathbb{F}_2^3$. For the description of the code let $\alpha$ be a primitive element in $\mathbb{F}_8$ such that $\alpha^3 + \alpha + 1 = 0$. The seven two-dimensional subspaces can be written as $\{0, \alpha^i, \alpha^{i+1}, \alpha^{i+3}\}$, $0 \leq i \leq 6$. The addition of two two-dimensional subspaces $\{0, \alpha^i, \alpha^{i+1}, \alpha^{i+3}\}$, $\{0, \alpha^j, \alpha^{j+1}, \alpha^{j+3}\}$, $0 \leq i < j \leq 6$ is $\{0, \alpha^i + \alpha^j, \alpha^{i+1} + \alpha^{j+1}, \alpha^{i+3} + \alpha^{j+3}\}$. It is readily verified that this definition produces a linear code.

**Remark 3.** There are three more different ways to define addition of the code $\Psi$. Again, we omit the details and leave them as an exercise to the reader. For each $n \geq 4$ there is a linear code $\mathcal{U}$ consisting of $2^{n-1} - 1$ two-dimensional subspaces, each one contains a given one-dimensional subspace $X$, and $2^{n-1}$ $(n-1)$-dimensional subspaces, each one does not contain the subspace $X$. The fact that $\mathcal{U}$ is linear is also left as an exercise to the reader.

For larger dimension we can construct linear codes having different dimension distribution than the ones we already mentioned by using a recursive construction with direct products of linear codes with smaller dimensions. We will omit the details, and leave them as an exercise to the reader, as this seems to be of less interest since the size of the linear codes is small as is the size of $\mathcal{L}$. Let $D_k^i$ denote the number of $k$-dimensional subspaces in a linear code $\mathcal{U}_i$. One can verify the following theorem.

**Theorem 7.** *If $D_0^1, D_1^1, \ldots, D_n^1$ is the dimension distribution of a linear code $\mathcal{U}_1$, over $\mathbb{F}_2^{n_1}$, and $D_0^2, D_1^2, \ldots, D_m^2$ is the dimension distribution of a linear code $\mathcal{U}_2$, over $\mathbb{F}_2^{n_2}$, then there is a recursive construction for a linear code $\mathcal{U}_3$, over $\mathbb{F}_2^{n_1+n_2}$, in which the number of codewords with dimension $\kappa$ is $\sum_{i=0}^{\kappa} D_i^1 \cdot D_{\kappa-i}^2$.*

Finally, the following theorem establishes a connection between complements and linearity.

**Theorem 8.** *Let $\mathcal{U}$ be a linear code in $\mathbb{P}_2(n)$, with $\mathbb{F}_2^n \in \mathcal{U}$, and let $\boxplus$ be the isometric linear addition on $\mathcal{U}$. Then the function $f : \mathcal{U} \to \mathcal{U}$ defined by $f(X) = \mathbb{F}_2^n \boxplus X$ is a complement on $\mathcal{U}$.*

PROOF. We have to prove that **P1** through **P4** are satisfied by the definition of $f$.
(i) By Definition 3 we have $d_S(X, \mathbb{F}_2^n \boxplus X) = d_S(X \boxplus X, X \boxplus (\mathbb{F}_2^n \boxplus X))$, and the equality $d_S(X \boxplus X, X \boxplus (\mathbb{F}_2^n \boxplus X)) = d_S(\{\mathbf{0}\}, \mathbb{F}_2^n) = n$ by Definition 2. Therefore, $X \cap (\mathbb{F}_2^n \boxplus X) = \{\mathbf{0}\}$ and $X \boxplus (\mathbb{F}_2^n \boxplus X) = \mathbb{F}_2^n$ and hence $f$ satisfies **P1**.
(ii) By Definition 3 we have $d_S(\mathbb{F}_2^n \boxplus X, \mathbb{F}_2^n \boxplus Y) = d_S(X, Y)$ and hence **P4** is satisfied.
(iii) $f(f(X)) = f(\mathbb{F}_2^n \boxplus X) = \mathbb{F}_2^n \boxplus (\mathbb{F}_2^n \boxplus X) = X$ and hence **P3** is satisfied.
(iv) **P2** is satisfied as a consequence from Lemma 12.
Thus, the theorem is proved. □

It follows from Theorem 8 that one can easily define a (full) complement on the linear code $\mathcal{L}$ constructed in Theorem 6. But, as we have already observed, $|\mathcal{L}| = 2^n$ is small compared to $|\mathbb{P}_2(n)|$.

## 4. Lattices, Complements, and Codes

In this section we describe a general approach to subspaces considering lattices in order to show how subspace codes can be regarded as the $q$-analog of binary block codes. The basic properties and additional material can be viewed in [2].

*4.1. General Description and Fundamental Properties*

A lattice $(L, \vee, \wedge, \leq)$ is a set $L$ together with join operator $\vee$, meet operator $\wedge$ and partial order relation $\leq$. For all $x, y \in L$ these operators correspond to each other by $x \leq y \Leftrightarrow x \vee y = y \Leftrightarrow x \wedge y = x$. We say that $y \in L$ covers $x \in L$ if $x \leq y$ and there is no $z \in L$ such that $x \leq z$ and $z \leq y$. A maximal chain between two elements $x$ and $y$ such that $x \leq y$ is a sequence $z_0 = x, z_1, \ldots, z_{\pi-1}, z_\pi = y$, such that $z_{i+1}$ covers $z_i$ for each $0 \leq i \leq \pi - 1$.

The lattice is bounded by 0 and 1 if 0 is the meet and 1 is the join of all lattice elements. In a bounded lattice where all maximal chains between the same elements have the same length a mapping $r : L \to \mathbb{N}_0$ which is called rank function can be defined with the following properties:

(i) $r(x) + r(y) = r(x \vee y) + r(x \wedge y)$ for all $x, y \in L$,
(ii) $r(y) = r(x) + 1$ for all $y \in L$ covering $x \in L$,
(iii) $r(0) = 0$.

It is well known [2] that every lattice admitting such a rank function is a metric lattice with a distance function:

$$d(x,y) := r(x \vee y) - r(x \wedge y) = r(x) + r(y) - 2r(x \wedge y) \qquad (2)$$

As example we consider the set of binary vectors $\{0,1\}^n$ of length $n$. This set forms a lattice, the so-called Boolean lattice, with the bitwise or-operation | as join-and-operation & as meet. In the following the corresponding partial order, defined by equivalence of join, meet and partial order, is abbreviated by $\preceq$. The lower bound is the all-zero vector $\mathbf{0}$ while the upper bound is the all-one vector $\mathbf{1}$. The rank function of this lattice is defined by the weight $\operatorname{wt} : \{0,1\}^n \to [0,n]$ which counts the number of non-zero components in a vector. The corresponding distance between $x = (x_1, \ldots, x_n)$ and $y = (y_1, \ldots, y_n)$ is given by

$$d_H(x,y) = \operatorname{wt}(x|y) - \operatorname{wt}(x\&y) = \operatorname{wt}(x) + \operatorname{wt}(y) - 2\operatorname{wt}(x\&y) = |\{i \mid x_i \neq y_i\}|,$$

which is nothing but the Hamming metric.

A second well known representation of the Boolean lattice can be obtained by subsets. If $[n] = \{1, \ldots, n\}$ denotes the canonical $n$-set, the *power set lattice* $\mathcal{P}(n) = \{X \mid X \subseteq [n]\}$ bijectively corresponds to $\{0,1\}^n$:

A subset $X \in \mathcal{P}(n)$ can be represented by its characteristic vector $\chi(X) = (x_1, \ldots, x_n)$, i.e. $x_i = 1$ if $i \in X$ and $x_i = 0$ otherwise. The join is the union $\cup$ and the meet is the intersection $\cap$ of sets. The partial order is the inclusion relation $\subseteq$. The lower, respectively upper bound, is the empty set $\varnothing$, respectively the whole set $[n]$. The rank of an element $X \in \mathcal{P}(n)$ is given by the cardinality $|X|$ and the metric by

$$d_P(X,Y) = |X \cup Y| - |X \cap Y| = |X| + |Y| - 2|X \cap Y| = |\Delta(X,Y)|,$$

which is the cardinality of the difference set $\Delta(X,Y) := (X \setminus Y) \cup (Y \setminus X)$ of $X, Y \in \mathcal{P}(n)$.

The following identities which show the one-to-one correspondence between the lattices $(\{0,1\}^n, |, \&, \preceq)$ and $(\mathcal{P}(n), \cup, \cap, \subseteq)$ are easily verified:

(i) $\chi(X \cup Y) = \chi(X)|\chi(Y)$
(ii) $\chi(X \cap Y) = \chi(X)\&\chi(Y)$
(iii) $|X| = \operatorname{wt}(\chi(X))$
(iv) $d_P(X,Y) = d_H(\chi(X), \chi(Y))$

The well known concept of $q$-analogs replaces subsets by subspaces of a vector space over a finite field and their orders by dimensions of the subspaces. In particular the $q$-analog of the power set lattice $(\mathcal{P}(n), \cup, \cap, \subseteq)$ is the *linear lattice* $(\mathbb{P}_q(n), +, \cap, \leq)$, i.e. the projective space $\mathbb{P}_q(n)$ together with the addition $+$ of vector spaces as join and the intersection $\cap$ as meet. The rank of $X \in \mathbb{P}_q(n)$ is obviously the dimension $\dim(X)$ and the distance in this case is defined by

$$d_S(X,Y) = \dim(X+Y) - \dim(X \cap Y) = \dim(X) + \dim(Y) - 2\dim(X \cap Y).$$

Table 1 summarizes all three lattices.

Table 1: boolean, power set, and linear lattice

| lattice | $\{0,1\}^n$ | $\xrightarrow{\text{as sets}}$ $\mathcal{P}(n)$ | $\xrightarrow{q\text{-analog}}$ $\mathbb{P}_q(n)$ |
|---|---|---|---|
| join | \| | $\cup$ | $+$ |
| meet | & | $\cap$ | $\cap$ |
| partial order | $\preceq$ | $\subseteq$ | $\leq$ |
| lower bound | $\mathbf{0}$ | $\varnothing$ | $\{\mathbf{0}\}$ |
| upper bound | $\mathbf{1}$ | $[n]$ | $\mathbb{F}_q^n$ |
| rank | $\text{wt}(\cdot)$ | $\|\cdot\|$ | $\dim(\cdot)$ |
| metric | $d_H$ | $d_P$ | $d_S$ |

Since a binary block code $C$ is a subset of $\{0,1\}^n$ and hence corresponds to a subset of $\mathcal{P}(n)$, a subspace code $\mathbb{C}$ which is a subset of $\mathbb{P}_q(n)$ can be interpreted as the $q$-analog of binary block code.

A natural goal in the theory of $q$-analogs is the generalization of definitions and results from sets to vector spaces. In our case we may consider further refinements and properties of binary block codes and adapt them to subspace codes.

For instance the $q$-analog of a binary constant weight code $C \subseteq \{0,1\}^n$, i.e. a code where all elements $x \in C$ have the same weight, is a subspace code $\mathbb{C} \subseteq \mathbb{P}_q(n)$ where all elements $X \in \mathbb{C}$ have the same dimension, so-called constant dimension subspace code.

Other concepts we may consider to be generalized to subspace codes are the so-called complements and linear code. In order to achieve this goal we first consider the properties of ordinary binary linear codes:

In the vector coding a binary block code $C \subseteq \{0,1\}^n$ is linear if $C$ is closed under the addition operation $\boxplus$, which is the bitwise XOR-operation. In order to obtain a general definition of this addition in the lattice case we express $\boxplus$ in terms of lattice operations. This leads to the definition of a complement. As already discussed in Section 2 the complement in the vector space is a mapping $f : \{0,1\}^n \to \{0,1\}^n$ which reverses each bit in a vector $x$ and it satisfies the properties $f(x)|x = \mathbf{1}$ and $f(x)\&x = \mathbf{0}$. Finally, we can write

$$x \boxplus y = (x \& f(y)) | (y \& f(x))$$

or equivalently in the notation of sets

$$X \boxplus Y = (X \cap f(Y)) \cup (Y \cap f(X))$$

where $f(X) = \{i \in [n] \mid i \notin X\}$ denotes the complement of the subset $X$.

4.2. Complements in Lattices

Let $(L, \vee, \wedge, \leq)$ be a bounded lattice with rank function $r$ and corresponding metric function $d$. In the remaining part of this section we investigate bijective mappings $f : L \to L$ having certain properties:

**Q1.** $x \leq y \Rightarrow f(x) \geq f(y)$

**Q2.** $f(x \vee y) = f(x) \wedge f(y)$
**Q3.** $f(x \wedge y) = f(x) \vee f(y)$
**Q4.** $r(f(x)) = r(1) - r(x)$
**Q5.** $d(f(x), f(y)) = d(x, y)$

A well known result [2, §3, Lemma 1-2] considering the first three conditions **Q1**, **Q2**, and **Q3** is given in the following lemma.

**Lemma 9.** *Let $(L, \vee, \wedge, \leq)$ be a bounded lattice with rank $r$ and let $f : L \to L$ be a bijective mapping. Then the conditions **Q1**, **Q2**, and **Q3** are equivalent.*

**Lemma 10.** *Let $(L, \vee, \wedge, \leq)$ be a bounded lattice with rank $r$ and let $f : L \to L$ be a bijective mapping. Then **Q1** implies **Q4**.*

PROOF. Consider a maximal chain $0 = x_0 \leq \cdots \leq x_{n-1} \leq x \leq x_{n+1} \leq \cdots \leq x_{n+m} = 1$ containing an arbitrary but fixed $x \in L$. Application of $f$ and **Q1** yields the order reversed chain $1 = f(0) = f(x_0) \geq \cdots \geq f(x_{n-1}) \geq f(x) \geq f(x_{n+1}) \geq \cdots \geq f(x_{n+m}) = f(1) = 0$. From the first chain we get $r(x) = n$ and $r(1) = n + m$ and from the second chain we obtain $r(f(x)) = m$ and hence $r(f(x)) = r(1) - r(x)$. □

**Lemma 11.** *Let $(L, \vee, \wedge, \leq)$ be a bounded lattice with rank $r$ and corresponding metric $d$ and let $f : L \to L$ be a bijective mapping. Then both conditions **Q4** and **Q5** together are equivalent to **Q1**.*

PROOF. (i) Assume first that **Q4** and **Q5** hold. Let $x, y \in L$ with $x \leq y$. Then, by (2) we obtain
$$d(x, y) = r(x \vee y) - r(x \wedge y) = r(y) - r(x) .$$
On the other hand by (2) and **Q4** we have
$$d(f(x), f(y)) = r(f(x)) + r(f(y)) - 2r(f(x) \wedge f(y)) = 2r(1) - r(x) - r(y) - 2r(f(x) \wedge f(y)) .$$
Since **Q5** holds, the last two equations implies that
$$r(y) - r(x) = 2r(1) - r(x) - r(y) - 2r(f(x) \wedge f(y)) .$$
Simplifying yields $r(f(x) \wedge f(y)) = r(1) - r(y)$ and hence by **Q4** we have $r(f(x) \wedge f(y)) = r(f(y))$. This condition together with $f(x) \wedge f(y) \leq f(y)$ (since $x \wedge y \leq y$ for any $x, y \in L$) implies $f(x) \wedge f(y) = f(y)$ which is equivalent to $f(x) \geq f(y)$, i.e. **Q1** holds.

(ii) Assume that **Q1** holds. **Q4** follows directly from Lemma 10. By (2) we have
$$d(f(x), f(y)) = r(f(x) \vee f(y)) - r(f(x) \wedge f(y)) .$$
By Lemma 9, **Q1**, **Q2**, and **Q3** are equivalent and hence
$$d(f(x), f(y)) = r(f(x \wedge y)) - r(f(x \vee y)) .$$
By **Q4** we have
$$d(f(x), f(y)) = r(1) - r(x \wedge y) - [r(1) - r(x \vee y)] = r(x \vee y) - r(x \wedge y) ,$$
and therefore by (2) we have $d(f(x), f(y)) = d(x, y)$, i.e. **Q5** holds.

Thus, **Q4** and **Q5** together are equivalent to **Q1**. □

*4.3. On the existence of Quasi-Complements*

Edwin Clark [4] proved the following result in case of the linear lattice (which is the projective space):

**Theorem 9.** *There is no function $f : \mathbb{P}_q(n) \to \mathbb{P}_q(n)$ which satisfies* **P1** *and* **Q1**.

**Theorem 10.** *There is no quasi-complement on $\mathbb{P}_q(n)$ satisfying properties* **P1**, **P2**, *and* **P4**.

PROOF. By Lemma 11, **P1**, **P2**, and **P4** with the rank function $r(X) = \dim(X)$ for each subspace $X$ of $\mathbb{P}_q(n)$ imply **P1** and **Q1**. The claim follows now by Theorem 9.

**Lemma 12.** *A quasi-complement function $f$ on a subset $\mathcal{U}$ of $\mathbb{P}_q(n)$ which satisfies* **P1** *and either* **P3** *or* **P4** *must satisfy* **P2** *too.*

PROOF. Assume that $f$ satisfies **P1**. If $f(X) = Y$ then clearly $\dim(X) = n - \dim(Y)$.

(i) If $f$ satisfies **P3** then $f(X) \neq f(Z)$ for $X \neq Z$ and clearly $f$ is a bijection. Therefore, $f$ satisfies **P2**.

(ii) If $f$ satisfies **P4** then for a given $k$, and two distinct elements $X, Y \in \mathbb{G}_q(n,k)$ we have $0 \neq d_S(X,Y) = d_S(f(X), f(Y))$ and hence $f(X) \neq f(Y)$. Therefore, $|\mathcal{U}_{n-k}| \geq |\mathcal{U}_k|$ and similarly $|\mathcal{U}_k| \geq |\mathcal{U}_{n-k}|$. Thus, the equality holds, $|\mathcal{U}_k| = |\mathcal{U}_{n-k}|$, and $f$ satisfies **P2**.

**Corollary 1.** *There is no quasi-complement on $\mathbb{P}_q(n)$ satisfying properties* **P1**, **P3**, *and* **P4**.

To conclude, we have given a complete answer to the existence question of a quasi-complement on $\mathbb{P}_q(n)$ satisfying any subset of the properties $\{\mathbf{P1}, \mathbf{P2}, \mathbf{P3}, \mathbf{P4}\}$. The complete answer is summarized in Table 2.

## 5. Conclusion and Open Problems

We have considered various problems related to linear codes and complements in projective space. The following three problems seem to be the most interesting ones for future research.

1. Prove that the size of the largest linear code in $\mathbb{P}_2(n)$ is $2^n$ or show a linear code in $\mathbb{P}_2(n)$ of size $2^{n+1}$.
2. A first step to solve the previous problem is to consider the same question when $\mathbb{F}_2^n$ is a codeword. Specifically, we suggest the following problem: given a linear code $\mathbb{C}$, in $\mathbb{P}_2(n)$, which contains $\mathbb{F}_2^n$ as a codeword, prove or disprove that the number of codewords with dimension $k$ in $\mathbb{C}$ is at most $\binom{n}{k}$.
3. Prove that the largest subset of $\mathbb{P}_q(n)$ on which a complement can be defined is the set $\mathcal{V}_q(n) = \{X \in \mathbb{P}_q(n) : X \cap X^\perp = \{\mathbf{0}\}\}$.

Table 2: Existence of quasi-complements

| properties | existence | reference |
|---|---|---|
| **P1** | Yes | Theorem 2 |
| **P2** | Yes | Theorem 1 |
| **P3** | Yes | Theorem 1 |
| **P4** | Yes | Theorem 1 |
| **P1**, **P2** | Yes | Theorem 2 |
| **P1**, **P3** | Yes (iff $n$ odd or $q$ odd) | Theorem 4 |
| **P1**, **P4** | No | Lemma 12 and Theorem 10 |
| **P2**, **P3** | Yes | Theorem 1 |
| **P2**, **P4** | Yes | Theorem 1 |
| **P3**, **P4** | Yes | Theorem 1 |
| **P1**, **P2**, **P3** | Yes (iff $n$ odd or $q$ odd) | Theorem 3 |
| **P1**, **P2**, **P4** | No | Theorem 10 |
| **P1**, **P3**, **P4** | No | Corollary 1 |
| **P2**, **P3**, **P4** | Yes | Theorem 1 |
| **P1**, **P2**, **P3**, **P4** | No | Corollary 1 |

# References

[1] R. Ahlswede, H. K. Aydinian, and L. H. Khachatrian, On perfect codes and related concepts, Designs, Codes Crypto., vol. 22, pp. 221-237, 2001.

[2] G. Birkhoff, Lattice Theory, vol. 25, Amer. Math. Soc. Colloquium Publications, revised edition, 1948.

[3] L. Chihara, On the zeros of the Askey-Wilson polynomials, with applications to coding theory, SIAM J. Math. Anal., vol. 18, pp. 191–207, 1987.

[4] W. E. Clark, Matching subspaces with complements in finite vector spaces, Bull. Inst. Combin. Appl., vol. 6, pp. 33–38, 1992.

[5] W. E. Clark and B. Shekhtman, Covering by complements of subspaces, II, Proceedings of the American Mathematical Society, vol. 125, pp. 251–254, 1997.

[6] W. E. Clark and B. Shekhtman, Covering by complements of subspaces, Linear and Multilinear Algebra, vol. 40, pp. 1–13, 1997.

[7] T. Etzion and N. Silberstein, Error-correcting codes in projective space via rank-metric codes and Ferrers diagrams, IEEE Trans. Inform. Theory, vol. IT-55, pp. 2909–2919, July 2009.

[8] T. Etzion and A. Vardy, Error-correcting codes in projective space, IEEE Trans. Inform. Theory, vol. IT-57, pp. 1165–1173, February 2011.

[9] M. Gadouleau and Z. Yan, Constant-rank codes and their connection to constant-dimension codes, IEEE Trans. Inform. Theory, vol. IT-56, pp. 3207–3216, July 2010.

[10] R. Koetter and F. R. Kschischang, Coding for errors and erasures in random network coding, IEEE Trans. Inform. Theory, vol. 54, pp. 3579–3591, August 2008.

[11] A. Kohnert and S. Kurz, Construction of large constant dimensiona codes with a prescribed minimum distance, Lecture Notes in Computer Science, vol. 5393, pp. 31–42, 2008.

[12] W. J. Martin and X. J. Zhu, Anticodes for the Grassman and bilinear forms graphs, Designs, Codes, Crypto., vol. 6, pp. 73–79, 1995.

[13] M. Q. Rieck, Totally isotropic subspaces, complementary subspaces, and generalized inverses, Linear Algebra and Its Applications, vol. 251, pp. 239–248, 1997.

[14] M. Schwartz and T. Etzion, Codes and anticodes in the Grassman graph, J. Combin. Theory, Series A, vol. 97, pp. 27–42, 2002.
[15] N. Sendrier, On the dimension of the hull, SIAM J. Discr. Math., vol. 10, pp. 282–293, 1997.
[16] D. Silva, F. R. Kschischang, and R. Koetter, A Rank-metric approach to error control in random network coding, IEEE Trans. Inform. Theory, vol. IT-54, pp. 3951–3967, September 2008.
[17] D. B. West, Introduction to Graph Theory, Prentice Hall, 1996.
[18] S. T. Xia and F. W. Fu, Johnson type bounds on constant dimension codes, Designs, Codes, Crypto., vol. 50, pp. 163–172, 2009.